\documentclass[aps,prd,nofootinbib,showpacs,superscriptaddress,12pt]{revtex4}
\usepackage{txfonts}
\usepackage{graphicx}
\usepackage{dcolumn}
\usepackage{bm}
\usepackage{amssymb}
\usepackage{latexsym}

\begin{document}

\title{\bf Holographic Dark Energy Model \\ Characterized by the Conformal-age-like Length}
\author{Zhuo-Peng~Huang}
\email[Electronic address: ]{zphuang@nudt.edu.cn}
\affiliation{Department of Physics, National University of Defense Technology, Hunan 410073, China}
\affiliation{State Key Laboratory of Theoretical Physics (SKLTP)\\
Kavli Institute for Theoretical Physics China (KITPC) \\
Institute of Theoretical Physics, Chinese Academy of Sciences, Beijing 100190, China}
\author{Yue-Liang~Wu}
\email[Electronic address: ]{ylwu@itp.ac.cn}
\affiliation{State Key Laboratory of Theoretical Physics (SKLTP)\\
Kavli Institute for Theoretical Physics China (KITPC) \\
Institute of Theoretical Physics, Chinese Academy of Sciences, Beijing 100190, China}

\date{\today}

\begin{abstract}

A holographic dark energy model characterized by the conformal-age-like length scale $L= \frac{1}{a^4(t)}\int_0^tdt'~a^3(t') $ is motivated from the four dimensional spacetime volume at cosmic time $t$ in the flat Friedmann-Robertson-Walker universe. It is shown that when the background constituent with constant equation of state $w_m$ dominates the universe in the early time, the fractional energy density of the dark energy scales as $\Omega_{de}\simeq \frac94(3+w_m)^2d^2a^2$ with the equation of state given by $w_{de}\simeq-\frac23 +w_m$.  The value of $w_m$ is taken to be $w_m\simeq-1$ during inflation, $w_m=\frac13$ in radiation-dominated epoch and $w_m=0$ in matter-dominated epoch respectively. When the model parameter $d$ takes the normal value at order one, the fractional density of dark energy is naturally negligible in the early universe, $\Omega_{de} \ll 1$ at $a \ll 1$.  With such an analytic feature, the model can be regarded as a single-parameter model like the $\Lambda$CDM model, so that the present fractional energy density $\Omega_{de}(a=1)$ can solely be determined by solving the differential equation of $\Omega_{de}$ once $d$ is given. We further extend the model to the general case in which both matter and radiation are present. The scenario involving possible interaction between the dark energy and the background constituent is also discussed.

\end{abstract}


\maketitle


\section{Introduction}

Observations\cite{Riess:1998cb, Perlmutter:1998np, Spergel:2003cb, Tegmark:2003ud} indicate that our universe has entered into a phase of accelerated expansion recently. Within the framework of the general relativity, the acceleration can phenomenally be attributed to the existence of a mysterious negative pressure component named as dark energy. The simplest candidate is a positive cosmological constant. Although fitting the observations well, a cosmological constant, however, is plagued with the fine-tuning problem and the coincidence problem\cite{Weinberg:1988cp}. A lot of alternative models (for some recent reviews see\cite{Copeland:2006wr, Li:2011sd}) have been proposed to explain such an exotic energy component. An
interesting kind of models based on the holographic principle\cite{'tHooft:1993gx, Susskind:1994vu, Cohen:1998zx} are holographic dark energy models
\cite{Li:2004rb, Cai:2007us, Wei:2007ty, Gao:2007ep}, in which the holographic dark energy density scales as
 \begin{equation}
    \rho_{de}\propto M_P^2L^{-2}~,
 \end{equation}
where $L$ is a characteristic length scale of the universe and $M_P$ is the reduced Planck constant.

Let us write the four dimensional spacetime volume at cosmic time $t$ of the flat Friedmann-Robertson-Walker~(FRW) universe in the following form
 \begin{equation}
    \int d^3x \int _0^t dt' \sqrt{-g}=\left(a^3(t)\int d^3x \right) \cdot a(t) \cdot \left(\frac{1}{a^4(t)}\int_0^t dt'~a^3(t')~\right),
 \end{equation}
where $a(t)$ is the scale factor of the universe. Since $a^3(t)\int d^3x $ is the physical space volume and $a(t)$ is the scale factor, then $
\frac{1}{a^4(t)}\int_0^tdt'~a^3(t') $ can be used as a conformal-age-like parameter of the universe. Taking such conformal-age-like parameter as the characteristic length scale $L$
\begin{equation}
    L=\frac{1}{a^4(t)}\int_0^t dt'~a^3(t')
 \end{equation}
we simply arrive at a conformal-age-like holographic dark energy (CHDE) model. Note that an alternative holographic dark energy model \cite{Gao:2011} was also motivated by the four dimensional spacetime volume at cosmic time $t$ of the flat FRW universe, where the characteristic length scale $\bar{r}_N$ was proposed to be determined by the equation $\frac{d \bar{r}_N}{dt}=1-H\bar{r}_N$, its solution is given by $ \bar{r}_N=\frac{1}{a(t)}(\int_0^tdt'~a(t')+ \delta)$ with $\delta$ a constant length scale. Obviously, its length scale is different from our proposed conformal-age-like length scale $L= \frac{1}{a^4(t)}\int_0^tdt'~a^3(t')$.

In this note, we are going to analyze the holographic dark energy model characterized by the conformal-age-like length scale
$L= \frac{1}{a^4(t)}\int_0^tdt'~a^3(t') $. We would like to address that the key point in our holographic dark energy model is that the conformal-age-like characteristic length scale includes the inflationary epoch, so that it is easily known how the initial dark energy density scales during the inflation. As a consequence, the dark energy and the radiation don't start off the same order of magnitude. Thus it can be shown that our holographic dark energy model is consistent with the inflationary universe and also the cosmological constraints. In Sec.\,II, we concentrate on the establishment of the CHDE model and make a detailed investigation on the CHDE model; In Sec.\,III, we further extend the CHDE model to the case in which both matter and radiation energy are present; In Sec.\,IV, we consider the possible interaction between the dark energy and the background  constituent; our conclusions and remarks are given in Sec.\,V. \\

\section{Model Building}

Taking the conformal-age-like parameter as a characteristic length scale of the universe
 \begin{equation}
    L=\frac{1}{a^4(t)}\int_0^t dt'~a^3(t')=\frac{1}{a^4(t)}\int_0^a a'^3\frac{da'}{H'a'}~,  \label{l}
 \end{equation}
where $H\equiv \dot{a}/a$ is the Hubble parameter with $``\cdot"$ being the derivative with respect to cosmic time $t$. When taking the characteristic length scale $L$ of the universe to define the holographic dark energy, we can parameterize the dark energy density as
 \begin{equation}
    \rho_{de}=3d^2 M_p^2 L^{-2}~, \label{rho}
 \end{equation}
where $d$ is a positive constant parameter. Correspondingly, the fractional energy density is given by
 \begin{equation}
    \Omega_{de}=\frac{\rho_{de}}{3M_p^2H^2}=\frac{d^2}{H^2L^2} ~. \label{frho}
 \end{equation}

Considering a flat Friedmann-Robertson-Walker~(FRW) universe containing such holographic dark energy and constituent with constant equation of state (EoS) $w_m$,
e.g. $w_m\simeq-1$ during inflation, $w_m=\frac13$ in radiation-dominated epoch and $w_m=0$ in matter-dominated epoch respectively,  we have Friedmann equation
 \begin{equation}
    3M_p^2H^2=\rho_m +\rho_{de}  ~,
 \end{equation}
or in fractional energy densities
 \begin{equation}
    \Omega_{de}+\Omega_{m}=1  ~,   \label{fri}
 \end{equation}
where $ \Omega_{m}=\frac{\rho_{m}}{3M_p^2H^2}$. If each energy component is conservative respectively , we have conservation equations
 \begin{equation}
    \dot{\rho}_{i}+3H(1+w_{i})\rho_{i}=0  ~ \label{ceq}
 \end{equation}
with $i=m,\, de$. In fact, when assuming that the total energy is conserved and the energy of matter is self-conservative, then the dark energy is also self-conservative. By using Eqs.(\ref{l}), (\ref{rho}), (\ref{frho}) and (\ref{ceq}), we get the EoS of the dark energy
 \begin{equation}
    w_{de}=-1 - \frac83 + \frac2{3d}\frac{\sqrt{\Omega_{de}}}a~.\label{wde}
 \end{equation}

The conservation of the constituent with constant $w_m$ results in the density $\rho_m=C_1 a^{-3(1+w_m)}$, where $C_1$ is a constant proportionality coefficient. Combining with the definition of fractional energy densities and the Friedmann equation, it is not difficult to get the relation
 \begin{equation}
    \frac1{Ha}= \frac1{ \scriptstyle \sqrt{\frac{C_1}{3M_p^2}}} \sqrt{a^{(1+3w_m)}(1-\Omega_{de})}~.\label{ha}
 \end{equation}
From Eqs.(\ref{l}) and (\ref{frho}), we have
 \begin{equation}
    \int_0^a a'^3\frac{da'}{H'a'}=\frac{a^5 d }{\sqrt{\Omega_{de}}Ha} ~.
 \end{equation}
Substituting Eq. (\ref{ha}) into above equation and taking derivative with respect to $a$ in both sides, we get the differential equation of motion for
$\Omega_{de}$
 \begin{equation}
    \frac{d\Omega_{de}}{da}=\frac{\Omega_{de}}{a}(1-\Omega_{de})\left(3(1+w_m)+8-\frac2d\frac{\sqrt{\Omega_{de}}}a \right)~. \label{ode}
 \end{equation}

Note that the energy density Eq.(\ref{rho}), the fractional energy density Eq.(\ref{frho}), the EoS Eq.(\ref{wde}) and the differential equation Eq.(\ref{ode}) are all invariant under the transformation $a \to \frac{a}{a_0}, ~ d \to d a_0$, where $a_0$ may be taken as the present scale factor of the universe (the subscript ''0'' always denotes the present value of the corresponding quantity). Performing such a transformation and setting $a_0=1$, all expressions keep the same form. So that the parameter $d$ is regarded to has absorbed a factor $a_0$ with setting $a_0=1$.

Accepting that the universe has successively experienced the inflation epoch during which $w_m\simeq-1$, the radiation-dominated epoch during which $w_m= \frac13$,
and the matter-dominated epoch during which $w_m= 0$ before entering into a phase of accelerated expansion recently, we can  approximately investigate the evolution
of $L$ and the fractional density $\Omega_{de}$ in the early universe. For self-consistency of the model, it is necessary to check whether the  fractional density
$\Omega_{de}$ in the early universe derived from directly calculation is consistent with the approximate solution of the differential equation Eq.(\ref{ode}) under
the limit $1-\Omega_{de} \simeq 1$ when $a \ll 1$.

When the constituent with constant $w_m$ dominates the universe, we have approximately $H^2 \propto a^{-3(1+w_m)}$ from Fridemann equations. Referring to Eq.(\ref{l}), we have
 \begin{equation}
    L=\left( \frac{a_i}{a} \right)^4 L_i+\frac{2}{3(3+w_m)}\left(\frac{1}{Ha}- \frac{1}{H_ia_i}\left( \frac{a_i}{a} \right)^4  \right)~,  \label{lm}
 \end{equation}
where subscript $i$ denotes the value of the corresponding quantity at the beginning of background constituent-dominated epoch. Particularly, we have
$L_i=\frac{1}{a_i^4}\int_0^{t_i} dt'~a^3(t')$.

During the epoch of inflation, the Hubble parameter $H$ is approximately a constant when the universe expands exponentially or $H\sim 1/t$ for a power-law inflation $a(t) \sim t^p$ with $p \sim 60$ \cite{YLW}. If the pre-inflation part $L_i$ is not very large unnaturally, those terms with factor $\left( \frac{a_i}{a} \right)^4$ in Eq.(\ref{lm}) can be ignored safely comparing to the term without such a factor. Therefore, in the inflation epoch we have $ L \simeq \frac{1}{3}\frac{1}{Ha}$. Thus the fractional energy density of dark energy is given by $\Omega_{de}\simeq 9d^2a^2$ from Eq.(\ref{frho}).

For the radiation-dominated epoch, $L_i\sim O(\frac{1}{H_ea_e})$ is set at the end of inflation.  when radiation dominates the universe, we have $H^2 \propto a^{-4}$, i.e., $\frac1{Ha}\propto a$. Therefore, those terms with factor $\left( \frac{a_i}{a} \right)^4$ in Eq.(\ref{lm}) can also be neglected safely comparing to the term without such a factor. Thus, in the radiation-dominated epoch we yield $ L\simeq \frac{1}{5}\frac{1}{Ha}$ and  $\Omega_{de}\simeq 25d^2a^2$. Similarly, we get $
L\simeq \frac{2}{9}\frac{1}{Ha}$ and $\Omega_{de}\simeq \frac{81}{4}d^2a^2$ in the matter-dominated epoch.

Summarily, when background constituent dominates the universe at early time, we have
 \begin{equation}
    L\simeq\frac{2}{3(3+w_m)}\frac{1}{Ha}~  \label{lm2}
 \end{equation}
and
 \begin{equation}
    \Omega_{de}\simeq\frac94(3+w_m)^2d^2a^2~.  \label{frhom}
 \end{equation}
The EoS $w_m$ of the background constituent are $w_m\simeq-1$ during inflation, $w_m=\frac13$ in radiation-dominated epoch and $w_m=0$ in matter-dominated epoch
respectively.

It is seen that when the parameter $d$ takes a normal value at order one, the fractional density of dark energy is natural small in the early time of universe $\Omega_{de} \ll 1$ as $a_i \ll 1$. Namely, the dark energy can be ignored in the early universe, which is consistent with the expansion history we have adopted.

It is not difficult to check that Eq.(\ref{frhom}) is an approximate solution of the differential equation (\ref{ode}) of motion for $\Omega_{de}$ under the limit $1-\Omega_{de} \simeq 1$ when $a \ll 1$. Therefore, the model is shown to be self-consistent. With such an analytic feature, we can use such an approximate solution $\Omega_{de}\simeq \frac94(3+w_m)^2d^2a_{\rm i}^2$ at certain point with $a_{\rm i}\ll 1$ as the initial condition to solve the differential equation of motion for $\Omega_{de}$. Notice that once $d$ is given, the present fractional energy density $\Omega_{de}(a=1)$ can be obtained by solving Eq.(\ref{ode}). In this sense, our present dark energy model is a single-parameter model like the $\Lambda$CDM model.

Substituting Eq.(\ref{frhom}) into Eq.(\ref{wde}), we get the EoS of the dark energy in background-constituent-dominated epoch
\begin{equation}
 w_{de}=-\frac23 +w_m
\end{equation}
Particularly, the EoS of dark energy is $-\frac53$ in the inflation epoch, then transits to $-\frac13$ in the radiation-dominated epoch, and further to $-\frac23$ in matter-dominated epoch. Referring to Eq.(\ref{wde}), the EoS of dark energy $w_{de}$ eventually transits back to below $-1$ due to the expansion of the universe. Such an CHDE model will be responsible of the cosmic accelerated expansion.

\section{A more general analysis on CHDE model}

Observations indicate that the universe mainly consists of dark energy, matter and radiation energy. For a more general analysis, we are going to consider a flat Friedmann-Robertson-Walker~(FRW) universe containing matter, radiation and CHDE. The corresponding Friedmann equation in fractional energy densities reads
 \begin{equation}
   \Omega_m+\Omega_r+\Omega_{de}=1 ~,  \label{fri2}
 \end{equation}
where the fractional energy densities are defined by $\Omega_i=\rho_i/\rho_c$ for $i=m$, $r$ and $de$, and $\rho_c=3M_p^2H^2$ is the critical energy density.

When there are no direct interchanges among different components, each energy component is conservative respectively. The EoS of dark energy is still given by Eq.(\ref{wde}). The conservations of matter and radiation result in $\rho_{m}=\rho_{m0}a^{-3}=\Omega_{m0} 3M_p^2H_0^2 a^{-3}$ and $\rho_{r}=\rho_{r0}a^{-4}=\Omega_{r0} 3M_p^2H_0^2
a^{-4}$ respectively. Defining $ r_0=\Omega_{r0} / \Omega_{m0} $, we have
\begin{equation}
\frac{r_0}{ a} = \frac{\Omega_r }{\Omega_{m}}
\end{equation}
Combining with the definition of fractional energy densities and the Friedmann equation, Eq.(\ref{ha}) is modified to be
 \begin{equation}
    \frac1{Ha}=\frac1{H_0\sqrt{\Omega_{m0}}}\sqrt{a(1-\Omega_{de})}\sqrt{\frac 1{1+r_0/a}}~.\label{ha2}
 \end{equation}
Following similar derivation in previous section, we get the differential equation of motion for $\Omega_{de}$
 \begin{equation}
    \frac{d\Omega_{de}}{da}=\frac{\Omega_{de}}{a}(1-\Omega_{de})\left(11+\frac{r_0/a}{1+r_0/a}-\frac2d\frac{\sqrt{\Omega_{de}}}a\right) ~. \label{ode2}
 \end{equation}
Under the limit $1-\Omega_{de} \simeq 1$ when $a \ll 1$, we get the following solution
 \begin{equation}
    \Omega_{de}\simeq\frac{d^2}{4}\left( 9+\frac{r_0/a}{1+r_0/a} \right)^2 a^2 ~ \label{frhom2}
 \end{equation}
which provides a good approximation for the differential equation (\ref{ode2}). With the approximate solution given in Eq.(\ref{frhom2}), it can easily be shown that the limits $r_0 \to 1$ and $r_0 \to 0$ consistently reduce to the cases discussed in previous section corresponding to $w_m=\frac13$ and $w_m=0$ respectively.

Taking the approximate solution $\Omega_{de}=\frac{d^2}{4}\left( 9+\frac{r_0/a}{1+r_0/a} \right)^2 a^2$ at the point $a \ll 1$ as the initial condition to solve the differential equation given in Eq.(\ref{ode2}), we can obtain the present fractional energy density $\Omega_{de}(a=1)$ once the parameter $d$ is given.

\section{Interaction with background constituent}

We may discuss the possible interaction between the CHDE and background constituent with a constant
$w_m$. Assuming that the CHDE exchanges energy with background constituent through an interaction term $Q$ as
 \begin{eqnarray}
    \dot{\rho}_{de}+3H\rho_{de}(1+w_{de})&=&-Q ~, \\
    \dot{\rho}_{m}~+3H\rho_{m}~(1+w_m) &=& ~Q ~,
 \end{eqnarray}
so that he total energy is still conservative $\dot{\rho}_{tot}+3H(\rho_{tot}+p_{tot})=0$. In this case, it is not difficult to find that the equation of motion for $\Omega_{de}$ is modified to be
 \begin{equation}
 \frac{d\Omega_{de}}{da}=\frac{\Omega_{de}}{a}\left[(1-\Omega_{de})\left(3(w_m+1)+8-\frac2d \frac{\sqrt{\Omega_{de}}}{a} \right)- \frac{Q}{3M_p^2H^3}\right] ~,\label{odeint}
 \end{equation}
and the EoS of holographic dark energy to be
 \begin{equation}
 w_{de}=-1-\frac83+\frac2{3d}\frac{\sqrt{\Omega_{de}}}{a}-\frac{Q}{3H\rho_{de}}~.\label{wdeint}
 \end{equation}
Obviously, Eqs.(\ref{odeint}) and (\ref{wdeint}) reduce to Eqs.(\ref{ode}) and (\ref{wde}) in the case of $Q=0$ (i.e. without interaction). \\

\section{Concluding remarks}

In this note, we have proposed a conformal-age-like holographic dark energy (CHDE) model characterized by the length scale $L= \frac{1}{a^4(t)}\int_0^tdt'~a^3(t')$ which is motivated from the four dimensional spacetime volume at cosmic time $t$ of the flat FRW universe. It has been shown that when the background constituent dominates the universe in the early time, the fractional energy density of the dark energy scales as $\Omega_{de}\simeq\frac94(3+w_m)^2d^2a^2$. The EoS $w_m$ of the background constituent are taken to be $w_m\simeq-1$ during inflation, $w_m=\frac13$ in radiation-dominated epoch and $w_m=0$ in matter-dominated epoch respectively. For the parameter $d$ to be a normal value at order one, the
fractional density of dark energy is naturally small $\Omega_{de} \ll 1$ for $a \ll 1$, namely the dark energy can be ignored in early universe. It has been shown that the present fractional energy density $\Omega_{de}(a=1)$ can solely be determined once $d$ is given, thus our present model is a single-parameter model like the $\Lambda$CDM model.

The EoS of the dark energy in the background-constituent-dominated epoch has been found to be $w_{de}=-\frac23 +w_m$ in the CHDE model. Particularly,  it is given to be $w_{de}\simeq -\frac53$ in the inflation epoch, then transits to $w_{de}=-\frac13$ in the radiation-dominated epoch, and further to $w_{de}=-\frac23$ in matter-dominated epoch. Eventually, $w_{de}$ turns back to below $-1$ due to the expansion of the universe, so that the dark energy is responsible of the cosmic accelerated expansion at present, which will be investigated in detail elsewhere. We have investigated a more general case with both matter and radiation presented in the CHDE model. We have also discussed the scenario in which the interaction between the dark energy and the background constituent is simply included.

\section*{Acknowledgements}
We would like to thank Miao Li for useful discussions. The author (Z.P.H) would like to thank M. Q. Huang and M. Zhong for their helpful support. This work is supported in part by the National Basic Research Program of China (973 Program) under Grants No. 2010CB833000; the National Nature Science Foundation of China (NSFC) under Grants No. 10975170, 10975184, 10947016.


\end{document}